# BER ANALYSIS OF ITERATIVE TURBO ENCODED MISO WIRELESS COMMUNICATION SYSTEM UNDER IMPLEMENTATION OF Q-OSTBC SCHEME

M. Hasnat Kabir[1], Shaikh Enayet Ullah[1], Mustari Zaman[1] and Md. Golam Rashed[2]

[1]Department of Information and Communication Engineering,
Rajshahi University, Rajshahi-6205, Bangladesh
hasnatkabir@yahoo.com
[2]Department of Electronics and Telecommunication Engineering (ETE)
Prime University, Dhaka-1216, Bangladesh.

## ABSTRACT

*In this paper, a comprehensive study has been made to evaluate the performance of a MISO wireless communication system. The 4-by-1 spatially multiplexed Turbo encoded system under investigation incorporates Quasi-orthogonal space-time block coding (Q-STBC) and ML signal detection schemes under QPSK, QAM, 16PSK and 16QAM digital modulations. The simulation results elucidate that a significant improvement of system performance is achieved in QAM modulation. The results are also indicative of noticeable system performance enhancement with increasing number of iterations in Turbo encoding/decoding scheme.*

## 1. INTRODUCTION

Multiple antennas at the transmitter or receiver are becoming very common in wireless system due to their diversity and capacity benefits. Systems with multiple antennas require channel models that characterized both spatial and temporal characteristics of the channel. It can be used to increase data rates through multiplexing or to improve performance through diversity. Space-time block coding (STBC) has been established over the past decade as an efficient and effective transmit diversity scheme to combat detrimental effects of wireless fading channels for MIMO/MISO wireless communication systems. With implementation of STBCs, full transmit antenna diversity can be achieved because of their simple decoding algorithm. Without channel state information (CSI) at the transmitter side, the STBCs can effectively combat channel fading in the wireless communication systems. One attractive approach to transmit diversity is Alamouti code which is an elegant and influential STBC design for two transmits- antenna and single receiver system [1]. Full-diversity transmission and rate one can be achieved by this code using two time slots for signals with complex constellations, which is employed in most current commercial wireless systems [2]. Tarokh et al has been presented a generalization form of Alamouti code using orthogonal design concept for systems with an arbitrary number of transmit antennas [3,4].

Orthogonal space-time block codes (O-STBCs) achieve full transmit diversity and allow independent single-complex symbol maximum-likelihood (ML) decoding, which make them appealing for multi-input multi-output (MIMO) communication systems. However, full-diversity rate-one O-STBCs do not exist for MIMO systems with more than two transmits antennas. For systems with four transmit antennas, the rate limitation of O-STBCs is overcome by quasi-orthogonal space-time block codes (QO-STBCs) at the expense of diversity loss and increased decoding complexity. Full-rate full diversity QO-STBCs are then obtained by rotating signal constellations for some transmitted symbols in the codeword [5-9]. The present study has been made merely to observe the impact on estimated BERs of the Q-OSTBC encoded 4x1 spatially multiplexed wireless communication system**.**

This paper is organized as: in section 2, related works have been discussed. We present the mathematical model and derive the ML detection procedure in section 3 whereas section 4 focuses the communication system model which show the block representation of quisi-orthogonal space time block encoded MISO wireless communication system with implemented iterative turbo encoding scheme. In section 5, we explain our findings i.e. results and discussions. At final section 6 concludes this research work.

## 2. RELATED WORKS

Tarokh et al., follow IEEE [3] introduce space-time block coding which is a new paradigm for communication over Rayleigh fading channels using multiple transmit antennas. The space time block encoded data has split into n streams which are simultaneously transmitted by n transmit antennas. Maximum likelihood decoding has preferred through decoupling of the signals transmitted from different antennas than joint detection. Space time block codes constructed by using the classical mathematical framework of orthogonal design only exist for few sporadic values of n.

There is a proposal of a novel class of space frequency and Space – Time Frequency block codes based on Quasi-Orthogonal design over a frequency selective Rayleigh fading channel in [10]. The three dimensional coding i.e. Quasi-Orthogonal Space – Time – Frequency coding is capable of achieving rate-one and exploiting all the spatial, multi-path and temporal diversity gains of offered channel. It has a benefit from a reduced maximum likelihood decoding complexity.

In [11], MIMO system allowed by the multiple antennas performs precoding, diversity coding and spatial multiplexing. The outcome of these MIMO techniques is higher data rate or longer transmits range without requiring additional bandwidth or transmits power. Alamouti's STBC used for 2 transmit antenna as well as orthogonal STBC has used for 3 and 4 transmit antennas for detailed study of diversity coding for MIMO system. After that performance has analyzed using BPSK, QPSK, 16-QAM and 64-QAM modulation schemes.

C.E. Sterian et. al., [12] have demonstrated a superquasi-orthogonal space time trellis code for four transmit antennas with rectangular signal constellations for wireless communication with a spectral efficiency of 4 bits/s/Hz.

## 3. MATHEMATICAL MODEL

In our present study, a wireless communication system is considered with four transmitting antennas at the base station (n=4) and a single antenna at the receiving end (m=1). The following assumptions are considered for the present cases:

(i) The communication channel is a flat fading

(ii) The path gain from transmit antenna to receive antenna is defined as $\alpha_{n,m}$.

(iii) The path gains are followed by the nature of independent complex Gaussian random variables.

(iv) The real and the imaginary parts of path gain have equal variance of 0.5. It can be relaxed without any change to the method of encoding and decoding.

(v) The wireless channel is quasi-static. Therefore, the path gains are constant over a frame of four-time slot with length T and vary from one frame to another.

The digitally modulated symbols are rearranged into blocks and each block contains four independent information symbols, $x_1, x_2, x_3$ and $x_4$ generated from the rectangular QAM/16QAM or circular QPSK/16PSK constellations. Under implementation of quasi-orthogonal STBC codes, the code matrix $C$ presented in Equation (1) is based on the orthogonal two-antenna Alamouti scheme and $x_i \in \Omega$ are the transmitted symbols taken from Ω. We generally use (* ) to denote complex conjugate . Each row of Equation (1) is transmitted in one time step and the received signal r after 4 consecutive time steps can be expressed as

$$C = \begin{bmatrix} x_1 & x_2 & x_3 & x_4 \\ -x_2^* & x_1^* & -x_4^* & x_3^* \\ -x_3^* & -x_4^* & x_1^* & x_2^* \\ x_4 & -x_3 & -x_2 & x_1 \end{bmatrix} \quad \textbf{(1)}$$

At the receiver side, for m antenna, the received signal $r_{t,m}$ at time t can be expressed as

$$r_{t,m} = \sum_{n=1}^{N} \alpha_{n,m} C_{tn} + \eta_{t,m} \quad \textbf{(2)}$$

where $\alpha_{n,m}$ is the path gain and $\eta_{t,m}$ are the noise samples which are independent complex Gaussian random variable with zero-mean. The real and the imaginary parts of noise have equal variance of N/(2 SNR). Symbols average energy which is transmitted from each antenna is normalized. Therefore, the average power of the received signal at each receive antenna is N with a signal-to-noise ratio of SNR.

Assuming perfect channel knowledge at the receiver, it computes the following decision metric

$$\sum_{m=1}^{M} \sum_{t=1}^{T} \left| r_{t,m} - \sum_{n=1}^{N} \alpha_{n,m} g_{tn} \right|^2 \quad \textbf{(3)}$$

over all possible $x_k = s_k \in C$ and decides in favor of the constellation symbols $s_{1,\bullet\bullet\bullet\bullet\bullet,} s_K$ that minimize the sum in maximum-likelihood decoding scheme; $g_{tn}$ is the complex space-time block coded transmission matrix with time t and the transmitting antenna n. according to [3], this metric is the sum of K components each involving only the variable $x_k, k = 1,2,......., K$. Due to the orthogonality, the sum has K components involving only the variable $x_k, k = 1,2,......., K$.

Using the orthogonality, the maximum-likelihood decision metric (3) can be calculated as the sum of two terms $f_{14}(x_1, x_4) + f_{23}(x_2, x_3)$, where $f_{14}$ is independent of $x_2$ and $x_3$ and $f_{23}$ is independent of $x_1$ and $x_4$. Thus, the minimization of (3) is equivalent to minimizing these two terms independently. In other words, first the decoder finds the pair $(s_1, s_4)$ that minimizes $f_{14}(x_1, x_4)$ among all possible $(x_1, x_4)$ pairs. Then, the decoder selects the pair $(s_2, s_3)$ which minimizes $f_{23}(x_2, x_3)$. This reduces the complexity of decoding without sacrificing the performance.

Simple manipulation of equation (3) provides the following formulas for $f_{14}(.)$ and $f_{23}(.)$:

$$f_{14}(x_1, x_4) = \sum_{m=1}^{M} \Bigg( \left( \sum_{n=1}^{4} |\alpha_{n,m}|^2 \right) \left( |x_1|^2 + |x_4|^2 \right)$$
$$+ 2\,\mathrm{Re}\{ (-\alpha_{1,m} r_{1,m}^{*} - \alpha_{2,m}^{*} r_{2,m} - \alpha_{3,m}^{*} r_{3,m} - \alpha_{4,m} r_{4,m}^{*}) x_1 + (-\alpha_{4,m} r_{1,m}^{*} + \alpha_{3,m}^{*} r_{2,m}$$
$$+ \alpha_{2,m}^{*} r_{3,m} - \alpha_{1,m} r_{4,m}^{*}) x_4 + (\alpha_{1,m} \alpha_{4,m}^{*} - \alpha_{2,m}^{*} \alpha_{3,m} - \alpha_{2,m} \alpha_{3,m}^{*} + \alpha_{1,m}^{*} \alpha_{4,m}) x_1 x_4^{*} \} \Bigg) \quad \textbf{(4)}$$

$$f_{23}(x_2, x_3) = \sum_{m=1}^{M} \Bigg( \left( \sum_{n=1}^{4} |\alpha_{n,m}|^2 \right) \left( |x_2|^2 + |x_3|^2 \right)$$
$$+ 2\,\mathrm{Re}\{ (-\alpha_{2,m} r_{1,m}^{*} + \alpha_{1,m}^{*} r_{2,m} - \alpha_{4,m}^{*} r_{3,m} + \alpha_{3,m} r_{4,m}^{*}) x_2 + (-\alpha_{3,m} r_{1,m}^{*} - \alpha_{4,m}^{*} r_{2,m}$$
$$+ \alpha_{1,m}^{*} r_{3,m} + \alpha_{2,m} r_{4,m}^{*}) x_3 + (\alpha_{2,m} \alpha_{3,m}^{*} - \alpha_{1,m}^{*} \alpha_{4,m} - \alpha_{1,m} \alpha_{4,m}^{*} + \alpha_{2,m}^{*} \alpha_{3,m}) x_2 x_3^{*} \} \Bigg) \quad \textbf{(5)}$$

where, Re is the real part of $\alpha$.

## 4. Communication System Model

A simulated single -user 4 x 1 spatially multiplexed wireless communication system as depicted in Figure 1 utilizes Quasi-Orthogonal space-time block and Turbo coding schemes. In such a communication system, the synthetically generated information bits are channel encoded using iterative-based Turbo encoding scheme and interleaved for minimization of burst errors. The interleaved bits are digitally modulated using various types of digital modulations such as Quadrature Phase Shift Keying (QPSK), Quadrature Amplitude (QAM), 16PSK and 16QAM [13,14]. The complex digitally modulated symbols are block encoded with implemented Q-OSTBC scheme and fed into four transmitting antennas. In receiving section, the transmitted signal is processed with ML decoding algorithm and the decoded modulated symbols are fed into Q-OSTBC decoder. Its output data are demapped, deinterleaved and eventually processed with Log-MAP decoding scheme to recover the transmitted data[15].

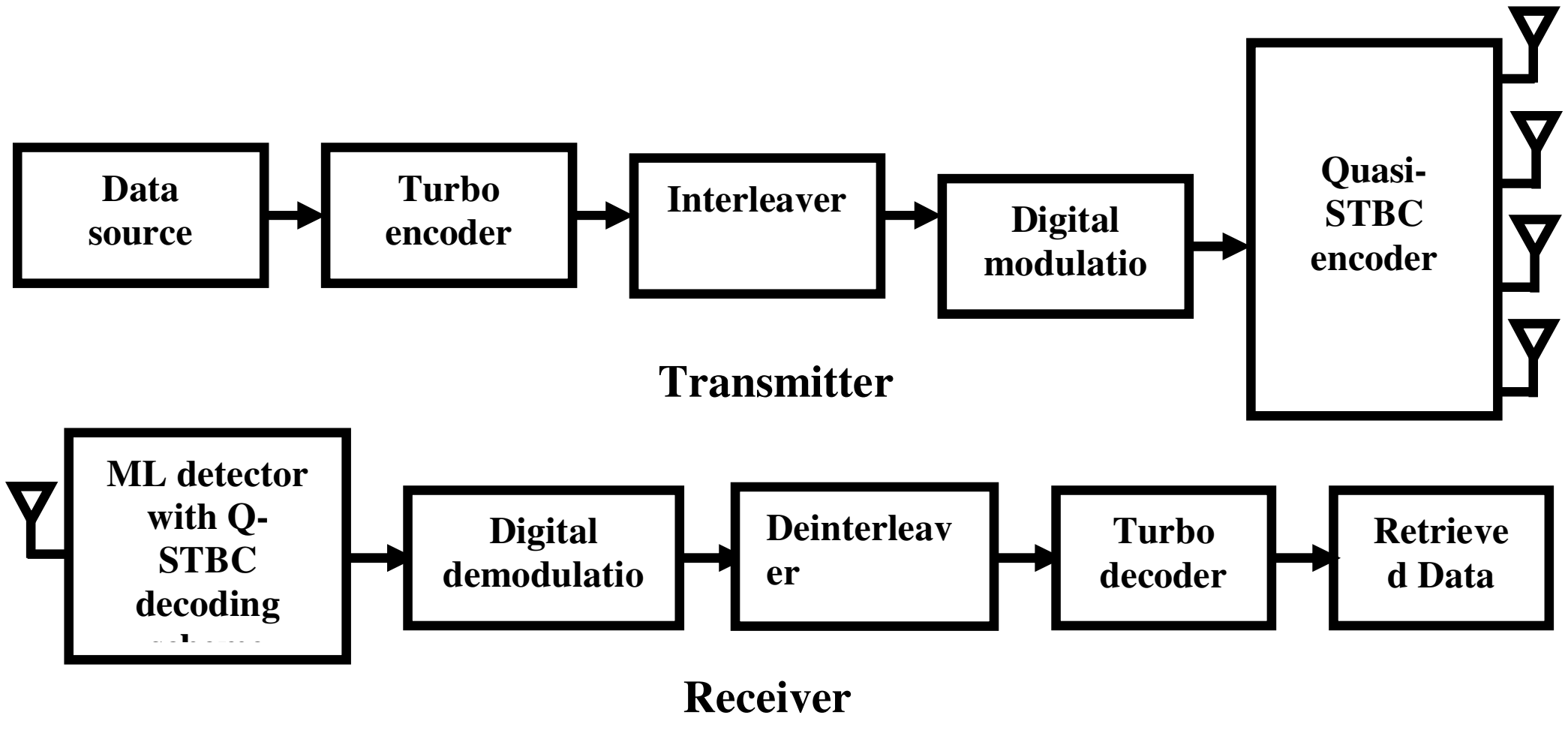

Figure 1: Block diagram of a Quasi-Orthogonal Space Time Block encoded MISO wireless communication system with implemented Iterative Turbo encoding scheme.

## 5. Results and Discussion

We have conducted computer simulations to evaluate the BER performance of a 4 x 1 spatially multiplexed Turbo encoded wireless communication system based on various parameters presented in Table 1. It is assumed that the channel state information (CSI) is available at the receiver and the fading process is approximately constant during each time slot assigned for simultaneous transmission of symbols from four transmitting antennas. The present study focuses the impact of Turbo encoding/decoding on the Q-OSTBC wireless communication system. The results are compared between different modulation techniques.

Table 1: Summary of the simulated model parameters

| Parameters | Values |
|---|---|
| No. of bits used | 1022 |
| Antenna configuration | 4-by-1 |
| Channel Coding /Decoding | Turbo coding/ Log-MAP decoding |
| Modulation | QPSK, QAM, 16PSK and 16QAM |
| Signal Detection Scheme | Maximum-Likelihood |
| Channel | AWGN and Rayleigh |
| Signal to noise ratio, SNR | 0 to10 dB |

The graphical illustrations presented in Figure 2 and Figure 3 show Q-OSTBC based system performance comparison without and with implementation of Turbo encoding scheme, respectively. In both cases, the system outperforms in QAM and shows worst performance in 16QAM digital modulations. The BER performance difference is quite obvious in lower SNR areas and the system's BER declines with increase in SNR values. In Figure 2, it is noticeable under channel uncoded situation that for a typically assumed SNR value of 2 dB, the BER values are **0.0069** and **0.5074** in case of QAM and 16QAM digital modulations viz., the system achieves a substantial gain of 18.67 dB in QAM as compared to 16QAM. In Figure 3, the BER values are **0.0055** and **0.3016** in case of QAM and 16QAM for a 2dB SNR value which is indicative of an enhancement of system performance by 17.39 dB. We have chosen QAM as the appropriate modulation schemes for the present study to illustrate the desired performance with implementation of the number of iteration of Turbo code. In Figure 4, it is observable that the BER performance of the system is improved with the increase in number of iterations in Turbo encoding scheme and the Bit error rate approaches zero at a comparatively low SNR value. In Figure 5, the transmitted and retrieved bits at a low SNR value of 2dB have been represented. The estimated bit error rate is 0.0039 and the retrieving capability of the iterative Turbo scheme based Quasi-Orthogonal Space Time Block encoded MISO wireless communication system is found to be quite satisfactory

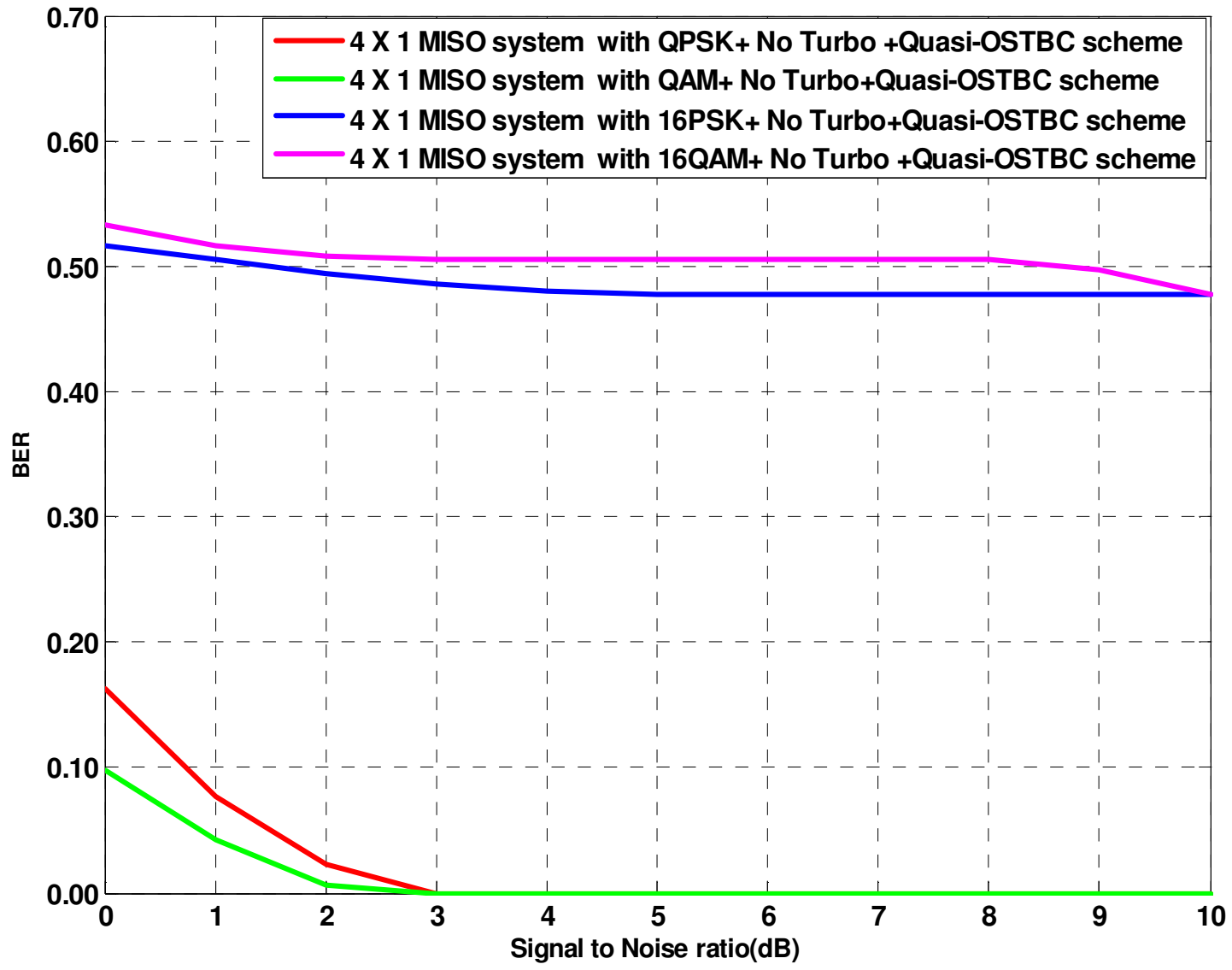

Figure 2: BER Performance of a Q-OSTB encoded MISO wireless communication system without implementation of Turbo encoding scheme

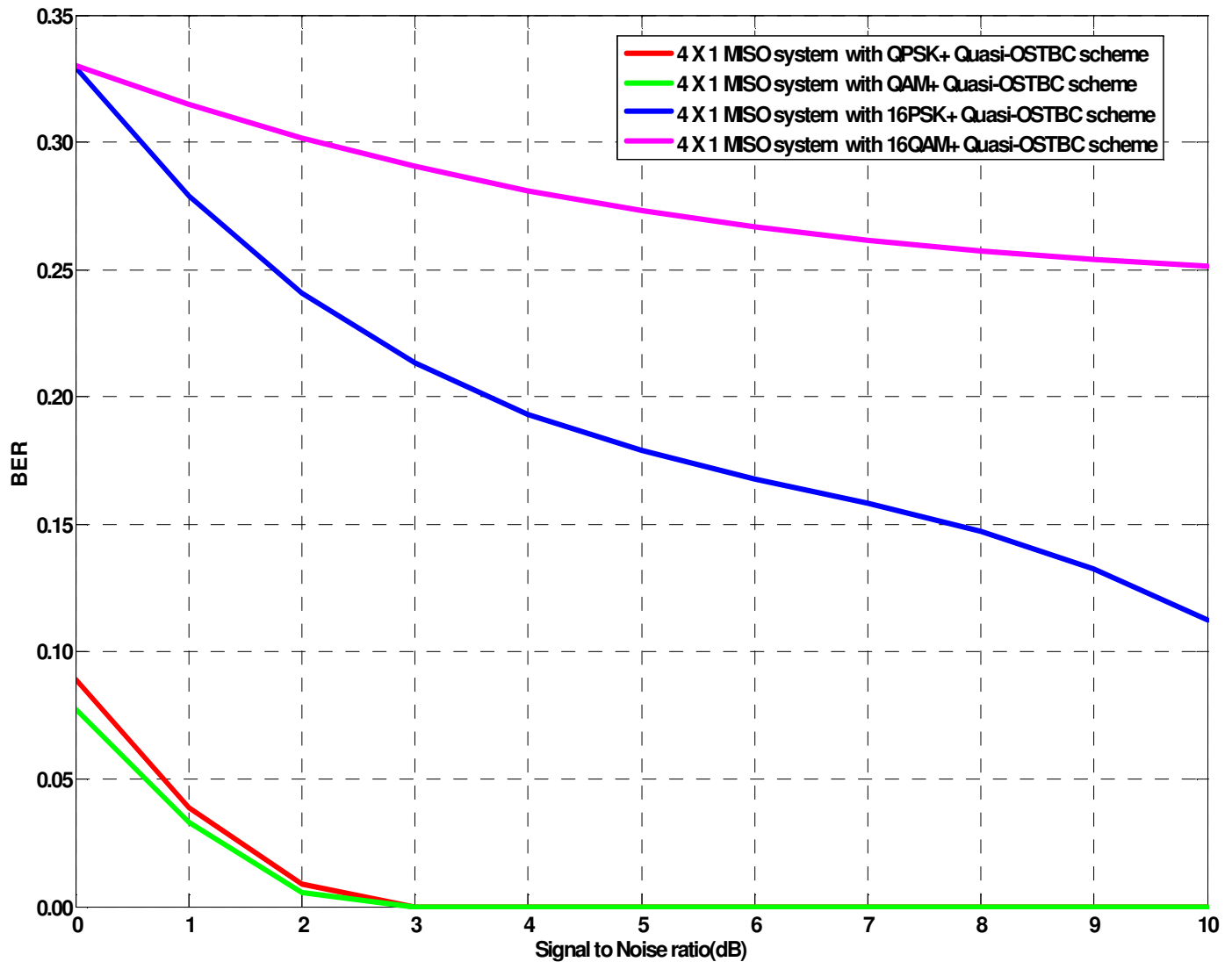

Figure 3: BER Performance of a Q-OSTB encoded MISO wireless communication system under implementation of Turbo encoding scheme

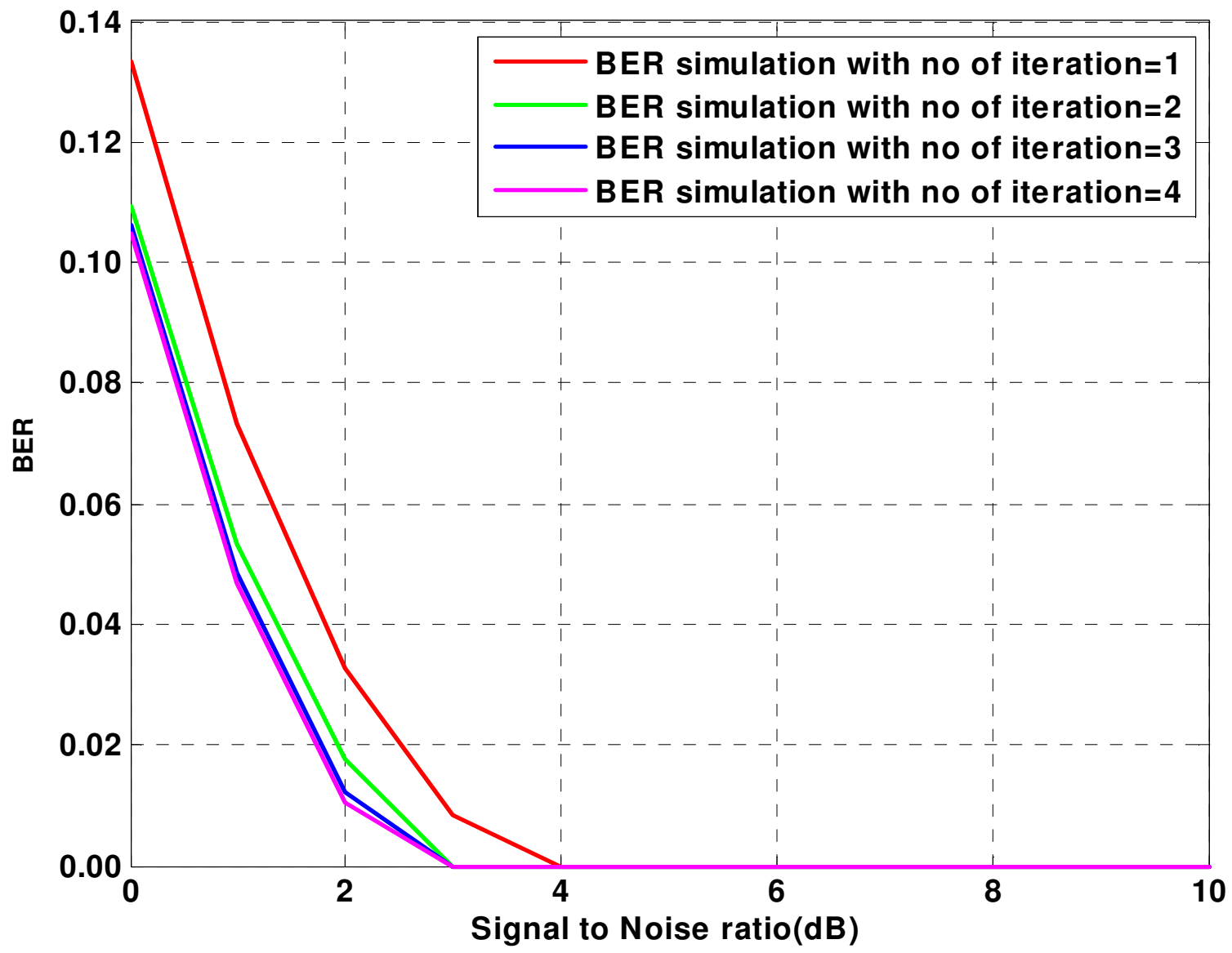

Figure 4: BER Performance of a Q-OSTB encoded MISO wireless communication system with different no of iterations under Turbo encoding and QAM digital modulation schemes.

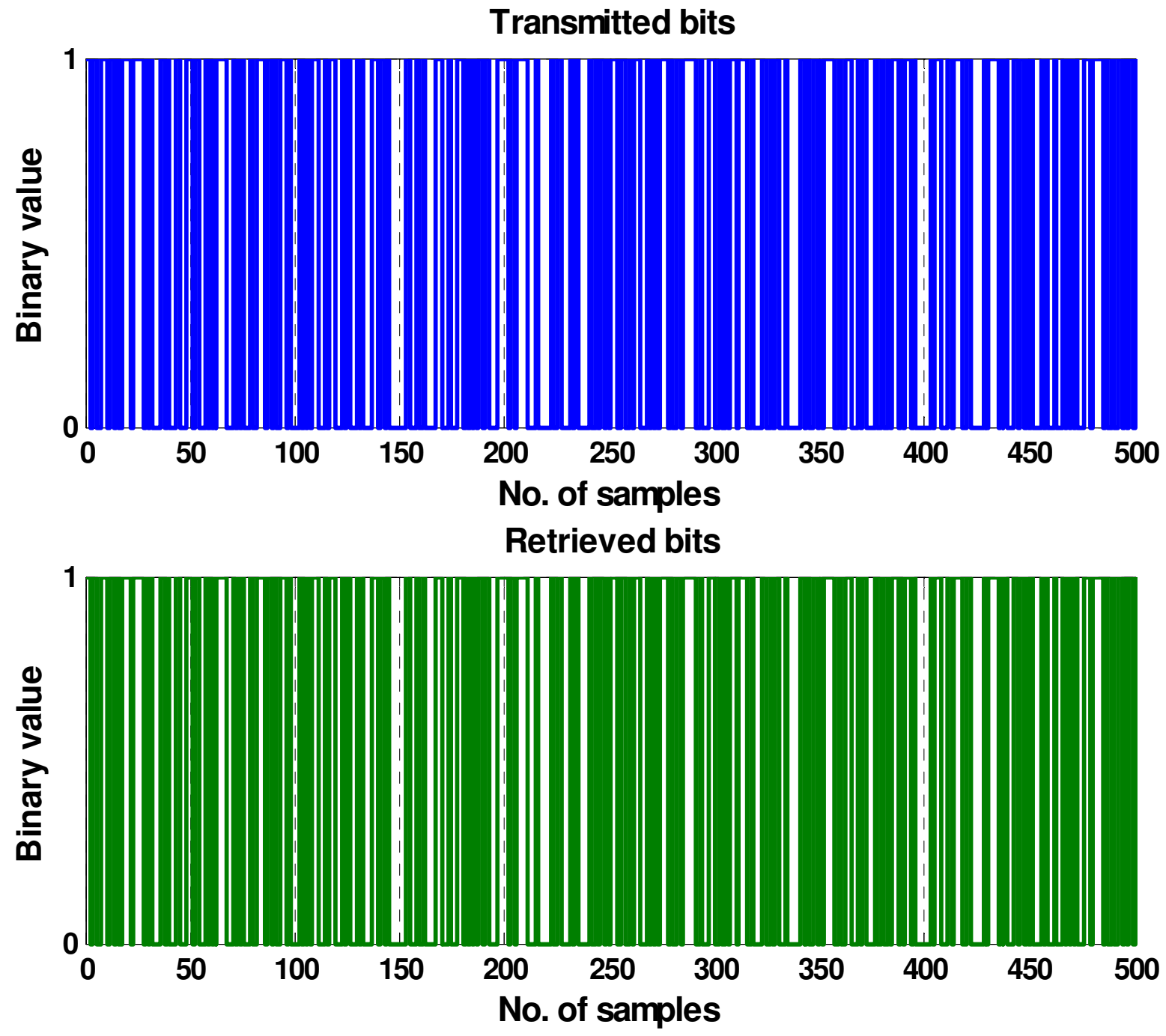

Figure 5: Transmitted and retrieved bits in a Q-OSTB encoded MISO wireless communication system under implementation of Turbo encoding scheme.

## 6. CONCLUSIONS

In this paper, we have presented simulation results concerning the adaptation of ML decoding algorithm in a Turbo encoded QOSTBC implementation based MISO wireless communication system. The system has used 4 transmit antennas and 1 received antenna. The performance has been investigated using four different modulation techniques namely QPSK, QAM, 16PSK and 16QAM. A range of system performance results highlights the impact of Turbo encoding/decoding and QOSTBC schemes on synthetically generated bit stream. In the context of system performance, it can be concluded that the implementation of QAM digital modulation technique in ML signal detection technique provides satisfactory result for such a MISO wireless communication system.